\documentclass[]{myaa} 

\usepackage{graphicx}

\usepackage {amssymb}

\usepackage {mathrsfs}

\usepackage{txfonts}
\idline{1}{1}
\begin{document}

\title{NLTE determination of the sodium abundance in a homogeneous 
sample of extremely metal-poor stars}

\author {
S.M. Andrievsky\inst{1,2}\and
M. Spite\inst{1}\and
S.A. Korotin\inst{2}\and
F. Spite\inst{1}\and
P. Bonifacio\inst{1,3}\and
R. Cayrel\inst{1}\and
\\
V. Hill\inst{1}\and
P. Fran\c cois\inst{1}
}

\offprints{M. Spite\\
           e-mail: Monique.Spite@obspm.fr}
\institute {
   GEPI, CNRS UMR 8111, Observatoire de Paris-Meudon, F-92125 Meudon
   Cedex, France
\and
   Department of Astronomy and Astronomical Observatory, Odessa
   National University, Shevchenko Park, 65014 Odessa, Ukraine
\and
   CIFIST Marie Curie Excellence Team}
\date{}

\authorrunning{Andrievsky et al.}
\titlerunning{NLTE determination of the Sodium abundance}

\abstract
{Abundance ratios in extremely metal-poor (EMP) stars are a good
indication of the chemical composition of the gas in the earliest
phases of the Galaxy evolution.  It had been found from an LTE
analysis that at low metallicity, and in contrast with most of the
other elements, the scatter of [Na/Fe] versus [Fe/H] was surprisingly
large and that, in giants, [Na/Fe] decreased with metallicity.}
{Since it is well known that the formation of sodium lines is very
sensitive to non-LTE effects, to firmly establish the behaviour of the
sodium abundance in the early Galaxy, we have used high quality
observations of a sample of EMP stars obtained with UVES at the
VLT, and we have taken into account the non-LTE line formation of
sodium.}
{The profiles of the two resonant sodium D lines (only these sodium
lines are detectable in the spectra of EMP stars) have been computed
in a sample of 54 EMP giants and turn-off stars (33 of them with ${\rm
[Fe/H]<-3.0}$) with a modified version of the code MULTI, and compared
to the observed spectra.  }
{With these new determinations in the range ${\rm -4 <[Fe/H]< -2.5}$,
both [Na/Fe] and [Na/Mg] are almost constant {\sl with a low scatter}.
In the turn-off stars and "unmixed" giants (located in the low RGB):
${\rm [Na/Fe] = -0.21 \pm0.13}$ or ${\rm [Na/Mg] = -0.45 \pm 0.16}$.
These values are in good agreement with the recent determinations of
[Na/Fe] and [Na/Mg] in nearby metal-poor stars.  Moreover we confirm
that all the sodium-rich stars are "mixed" stars (i.e., giants located
after the bump, which have undergone an extra mixing).  None of the
turn-off stars is sodium-rich.  As a consequence it is probable that
the sodium enhancement observed in some mixed giants is the result of
a deep mixing.}
{}

\maketitle

\section{Introduction}

During the last decade the interest in the extremely metal-poor (EMP)
galactic stars has significantly increased.  These stars, in fact, are
reservoirs of the primordial Galactic material that was only slightly
polluted by nuclides produced in SNe~II (by the first generation of
massive stars). Thus, elemental abundances provided by
metal-poor old stars, and especially abundance ratios, are of the
highest importance for the testing of the SN theoretical models, their
yields, and finally for constraining the initial mass function slope
in the early Galaxy and understanding its formation history and early
evolution.

With the advent of large aperture telescopes, the precise elemental
abundances in these faint objects became available.  Such abundance
ratios as [$\alpha$/Fe], [Na, Al/Fe], and others can be derived from
high-resolution spectra and then compared to the SNe~II yield
prediction.  In particular in Cayrel et al.  (2004) and in Bonifacio et
al.  (2006), a homogeneous sample of EMP stars has
been studied: 35 giants and 18 turn-off stars (39 of them with ${\rm
[Fe/H] \leq -3}$).  It has been shown that in these EMP stars the scatter of the abundanc
e ratios at metallicity
below --3.0 dex is generally very small and comparable to the
measurement errors.  But there are some exceptions, and one of the
most striking cases is [Na/Fe], which varies from star to star by a
factor close to ten.
In these EMP stars, only the resonance lines of Na (D1 and D2) are
visible, and these lines are known to be very sensitive to NLTE
effects (e.g., Mashonkina et al.  1993; Baum\"uller et al. 1998).

In Cayrel et al.  (2004) and Spite et al.  (2005) the sodium abundance
has been calculated under the LTE hypothesis.  Later, as a very first
approximation, a uniform correction LTE-NLTE of --0.5 dex,
extrapolated in the table of Baum\"uller et al.  (1998) for $\log~g =
2$, was applied to all the giants.  The scatter of the relation
[Na/Fe] versus [Fe/H] in giants was then very large, and [Na/Fe] was
decreasing with the metallicity in contrast to the turn-off stars (Spite
et al., 2005) where the ratio [Na/Fe] was almost constant in the
interval ${\rm -3.5 < [Fe/H] < -2.5}$.

\begin{figure}
\begin {center}
\resizebox  {8.0cm}{5.0 cm} 
{\includegraphics{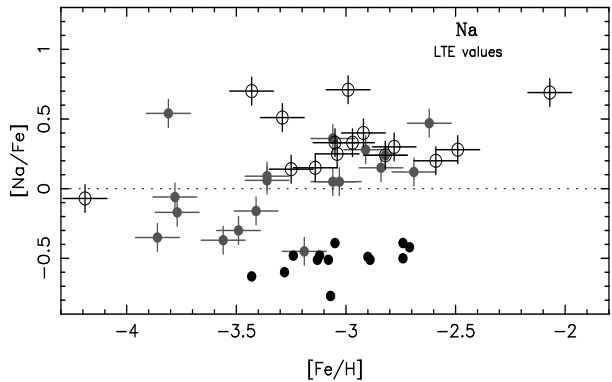}}
\caption[]{[Na/Fe] vs. [Fe/H] for an homogeneous sample of EMP stars
(Cayrel et al. 2004; Bonifacio et al. 2006): dwarfs (black dots),
low RGB stars (grey dots), and upper RGB stars (open circles).  The
abundance of Na is deduced from the D lines through an  LTE analysis.
As a mean, the ratio [Na/Fe] is lower in turn-off stars (black dots)
than in RGB stars and is also lower in lower RGB stars (grey dots)
than in upper RGB stars (open circles).  This graph suggests a
variation of the computed sodium abundance with the gravity, which
could be the consequence of an NLTE effect.
}
\label{na-lte}
\end {center}
\end{figure}

In Fig. \ref{na-lte} we show the relation [Na/Fe] vs. [Fe/H] for our
sample of dwarfs and giants without any NLTE correction.  The lower
RGB stars (below the bump, see Spite et al.  2006) and the upper RGB
stars (above the bump) have been plotted with different symbols.  In
the mean, the dwarfs have a lower value of [Na/Fe] than the giants,
and the lower RGB stars a lower value of [Na/Fe] than the upper RGB
stars.  This variation of [Na/Fe] with the gravity of the star
suggests that, as expected, non-LTE effects are very pronounced on the
Na D lines, and that as a consequence it is important to carefully take them
into account.
In this paper we present new determinations of the sodium abundance in
the sample of EMP stars of Cayrel et al.  (2004) and Bonifacio et al.
(2006) through a careful NLTE analysis, and we discuss the new trends
obtained.

\section{The star sample}

The spectra used here have been presented in detail in Cayrel et al.
(2004) and Bonifacio et al.  (2006).  The observations were performed
with the ESO VLT and its high resolution spectrograph UVES (Dekker et
al.  2000).  The spectra have a resolving power of 43000 in the region
of the D lines and a typical S/N ratio per pixel of $\sim 200$; they
have been reduced using the UVES context (Ballester et al.  2000)
within MIDAS. Stellar parameters (effective temperature, surface
gravity, microturbulent velocity, and metallicity) for each star have
been taken from Cayrel et al.  (2004) and Bonifacio et al.  (2006).

\section{Method of analysis} \label{analysis}

To derive the sodium abundance in the program stars, we perform an
NLTE analysis of the two resonant sodium D1 and D2 lines (only these
lines are seen in the spectra of the EMP stars) using a modified
version of the MULTI code (Carlsson 1986).  These modifications are
described in Korotin et al.  (1999ab).  To more adequately calculate
the continuous opacity taking into account the absorption produced by
the great number of spectral lines, the additional opacity sources
from ATLAS9 (Kurucz 1992) are included.  Simultaneous solution of the
radiative transfer and statistical equilibrium equations are obtained
using the approximation of a complete frequency redistribution for all
the sodium lines.  Atmosphere models of appropriate metallicity are
interpolated in Kurucz's grid of stellar atmosphere models
(microturbulent velocity 2~km~s$^{-1}$, $\alpha = 1.25$).  This
modified code has already been used for NLTE abundance determination
of carbon, oxygen, and sodium in stars of different types (see, e.g.,
Korotin et al.  1999ab; Korotin \& Mishenina 1999; Andrievsky et al.
2001; Andrievsky et al.  2002, and references therein).

We employ the modified model of the sodium atom (first considered by
Sakhibullin 1987) that consists of 27 energy levels of Na~I atoms in
addition to the ground level of the Na~II ion.
The fine splitting has been taken into account only for the 3p level.
This enables one to calculate the sodium doublet more precisely.
The radiative
transitions between the first 20 levels of Na~I and the ground level
of Na~II are considered, while transitions between the other levels
are used for the particle number conservation.  Linearization includes
46 $b-b$ and 20 $b-f$ transitions. Radiative rates for 34 transitions
are fixed.
Photoionization rates were taken from TOPBASE. Collisions with electrons,
as well as with hydrogen atoms, have been included.

\begin{figure}
\resizebox{\hsize}{!}{\includegraphics{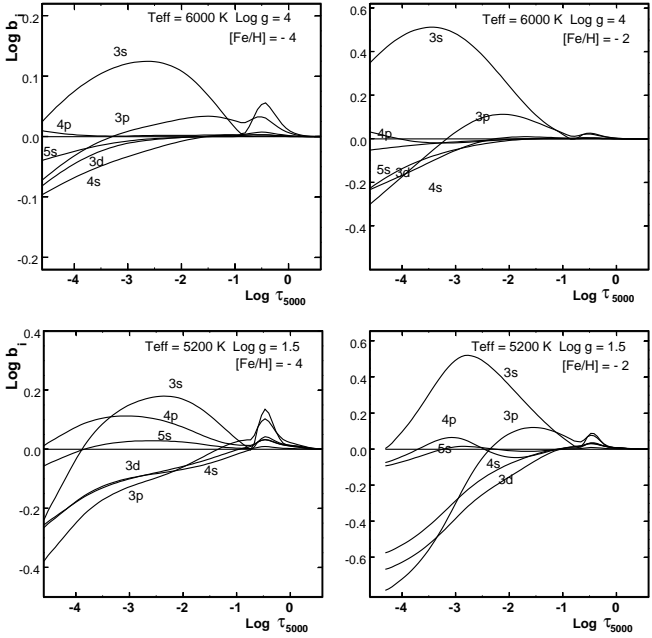}}
\caption[]{Departure coefficients in atmospheres of dwarfs and giants
of the different metallicities.}
\label{DepCoef}
\end{figure}

Figure  \ref{DepCoef} shows the behaviour of the departure coefficients
of Na~I for dwarfs and giants of different metallicities.  It is
clearly seen that within the region of the line-formation in the
atmosphere, the lower levels of the Na~I atom are overpopulated, leading
to a larger opacity in D1,2 lines compared to the pure LTE case, and
therefore to larger equivalent widths of these lines.  This means that
NLTE abundance corrections will be negative.

To find the relative-to-solar sodium abundance, we computed NLTE sodium
abundance in the Sun.  This was done with the following lines:
4496.05, 4982.81, 5148.84, 5682.63, 5688.20, 5889.95, 5895.92,
6154.22, 6160.75, 8183.25, 8194.82, and 12679.14 \AA. Their profiles
were extracted from the Kurucz et al.  (1984) solar flux spectrum.
Similarly to the stellar spectra analysis,  to derive the NLTE
sodium abundance in the Sun, we used the solar atmosphere model from the
Kurucz's grid and the solar microturbulent velocity recommended by Maltby et al.
(1986).
Our NLTE solar sodium abundance is ${\rm
\epsilon(Na)_{\odot}=6.25 \pm 0.04}$, in agreement with the value
found by Baum\"uller et al.  (\cite{BBG98}):
${\rm\epsilon(Na)_{\odot}=6.30 \pm 0.03}$, and Mashonkina et al.
(\cite{MET00}): ${\rm\epsilon(Na)_\odot}=6.20$.  The value ${\rm
\epsilon(Na)_{\odot}=6.25}$ will be used later, as a reference to
determine [Na/H].

\begin{table*}
\begin{center}    
\caption[]{Model parameters and NLTE sodium abundance in our sample of stars. 
The three stars marked with an asterisk are carbon-rich stars. An "m" 
in the last column denotes the mixed giants following Spite et al. (2005).}
\label{tabstars}
\begin{tabular}{lccccccccccc}
\hline
Star      & T$_{\rm eff},\,K$ & $\log~g$ & v$_{\rm t}$, km~s$^{-1}$ &
[Fe/H] & $\epsilon$(Na)& [Na/H] & [Na/Fe] &Rem\\
\hline
\\
Turnoff stars \\
\hline
BS~16023--046  & 6360 & 4.5 & 1.4 &  --2.97  &  3.33 &  --2.92 &   +0.05 & \\
BS~16968--061  & 6040 & 3.8 & 1.5 &  --3.05  &  2.80 &  --3.45 &  --0.40 & \\
BS~17570--063  & 6240 & 4.8 & 0.5 &  --2.92  &  3.12 &  --3.13 &  --0.21 & \\
CS~22177--009  & 6260 & 4.5 & 1.2 &  --3.10  &  2.95 &  --3.30 &  --0.20 & \\
CS~22888--031  & 6150 & 5.0 & 0.5 &  --3.28  &  2.70 &  --3.55 &  --0.25 & \\
CS~22948--093  & 6360 & 4.3 & 1.2 &  --3.43  &  2.44 &  --3.81 &  --0.51 & \\
CS~22953--037  & 6360 & 4.3 & 1.4 &  --2.89  &  2.90 &  --3.35 &  --0.46 & \\
CS~22965--054  & 6090 & 3.8 & 1.4 &  --3.04  &  3.10 &  --3.15 &  --0.11 & \\
CS~22966--011  & 6200 & 4.8 & 1.1 &  --3.07  &  2.95 &  --3.30 &  --0.23 & \\
CS~29499--060  & 6320 & 4.0 & 1.5 &  --2.70  &  3.23 &  --3.02 &  --0.32 & \\
CS~29506--007  & 6270 & 4.0 & 1.7 &  --2.91  &  3.15 &  --3.10 &  --0.19 & \\
CS~29506--090  & 6300 & 4.3 & 1.4 &  --2.83  &  3.40 &  --2.85 &  --0.02 & \\
CS~29518--020  & 6240 & 4.5 & 1.7 &  --2.77  &  3.20 &  --3.05 &  --0.25 & \\
CS~29518--043  & 6430 & 4.3 & 1.3 &  --3.24  &  2.75 &  --3.50 &  --0.30 & \\
CS~29527--015  & 6240 & 4.0 & 1.6 &  --3.55  &  2.50 &  --3.75 &  --0.20 & \\
CS~29528--041* & 6170 & 4.0 & 1.3 &  --3.06  &  3.85 &  --2.40 &   +0.66 & \\
CS~30301--024  & 6330 & 4.0 & 1.6 &  --2.75  &  3.30 &  --2.95 &  --0.20 &\\
CS~30339--069  & 6240 & 4.0 & 1.3 &  --3.08  &  3.00 &  --3.25 &  --0.17 &\\
CS~31061--032  & 6410 & 4.3 & 1.4 &  --2.58  &  3.50 &  --2.75 &  --0.17 &\\
\hline
\\
Giants  \\
\hline
HD~2796           & 4950 & 1.5 & 2.1 &  --2.47 &   3.65 & --2.60 & --0.13  & m\\
HD~122563       & 4600 & 1.1 & 2.0 &  --2.82 &   3.20 & --3.05 & --0.23  & m\\
HD~186478       & 4700 & 1.3 & 2.0 &  --2.59 &   3.45 & --2.80 & --0.21  & m\\
BD~+17:3248    & 5250 & 1.4 & 1.5 &  --2.07 &   4.44 & --1.81 &  +0.26  & m\\
BD~--18:5550    & 4750 & 1.4 & 1.8 &  --3.06 &   2.90 & --3.35 & --0.29  & \\
CD~--38:245      & 4800 & 1.5 & 2.2 &  --4.19 &   2.15 & --4.10 &  +0.09  & m\\
BS~16467--062  & 5200 & 2.5 & 1.6 &  --3.77 &   2.38 & --3.87 & --0.10  & \\
BS~16477--003  & 4900 & 1.7 & 1.8 &  --3.36 &   2.80 & --3.45 & --0.09  & \\
BS~17569--049  & 4700 & 1.2 & 1.9 &  --2.88 &   3.25 & --3.00 & --0.12  & m\\
CS~22169--035  & 4700 & 1.2 & 2.2 &  --3.04 &   3.30 & --2.95 &  +0.09  & m\\
CS~22172--002  & 4800 & 1.3 & 2.2 &  --3.86 &   2.15 & --4.10 & --0.24  & \\
CS~22186--025  & 4900 & 1.5 & 2.0 &  --3.00 &   3.15 & --3.10 & --0.10  & m\\
CS~22189--009  & 4900 & 1.7 & 1.9 &  --3.49 &   2.47 & --3.78 & --0.29  & \\
CS~22873--055  & 4550 & 0.7 & 2.2 &  --2.99 &   3.40 & --2.85 &  +0.14  & m\\
CS~22873--166  & 4550 & 0.9 & 2.1 &  --2.97 &   3.20 & --3.05 & --0.08  & m\\
CS~22878--101  & 4800 & 1.3 & 2.0 &  --3.25 &   2.95 & --3.30 & --0.05  & m\\
CS~22885--096  & 5050 & 2.6 & 1.8 &  --3.78 &   2.50 & --3.75 &  +0.03  & \\
CS~22891--209  & 4700 & 1.0 & 2.1 &  --3.29 &   3.10 & --3.15 &  +0.14  & m\\
CS~22892--052* & 4850 & 1.6 & 1.9 &  --3.03 &   3.00 & --3.25 & --0.22  & \\
CS~22896--154  & 5250 & 2.7 & 1.2 &  --2.69 &   3.47 & --2.78 & --0.09  & \\
CS~22897--008  & 4900 & 1.7 & 2.0 &  --3.41 &   2.69 & --3.56 & --0.15  & \\
CS~22948--066  & 5100 & 1.8 & 2.0 &  --3.14 &   2.97 & --3.28 & --0.14  & m\\
CS~22949--037* & 4900 & 1.5 & 1.8 &  --3.97 &   3.85 & --2.40 &  +1.57  & m\\
CS~22952--015  & 4800 & 1.3 & 2.1 &  --3.43 &   3.33 & --2.92 &  +0.51  & m\\
CS~22953--003  & 5100 & 2.3 & 1.7 &  --2.84 &   3.20 & --3.05 & --0.21  & \\
CS~22956--050  & 4900 & 1.7 & 1.8 &  --3.33 &   4.30 & --1.95 & - -     & \\
CS~22966--057  & 5300 & 2.2 & 1.4 &  --2.62 &   3.60 & --2.65 & --0.03  & \\
CS~22968--014  & 4850 & 1.7 & 1.9 &  --3.56 &   2.35 & --3.90 & --0.34  & \\
CS~29491--053  & 4700 & 1.3 & 2.0 &  --3.04 &   3.05 & --3.20 & --0.16  & m\\
CS~29495--041  & 4800 & 1.5 & 1.8 &  --2.82 &   3.22 & --3.03 & --0.21  & \\
CS~29502--042  & 5100 & 2.5 & 1.5 &  --3.19 &   2.63 & --3.62 & --0.43  & \\
CS~29516--024  & 4650 & 1.2 & 1.7 &  --3.06 &   3.05 & --3.20 & --0.14  & \\
CS~29518--051  & 5200 & 2.6 & 1.4 &  --2.69 &   3.40 & --2.85 & --0.16  & m\\
CS~30325--094  & 4950 & 2.0 & 1.5 &  --3.30 &   2.84 & --3.41 & --0.11  & \\
CS~31082--001  & 4825 & 1.5 & 1.8 &  --2.91 &   3.30 & --2.95 & --0.04  & \\
\hline
\end{tabular}
\end{center}
\end{table*}

\begin{figure}
\resizebox{\hsize}{!}{\includegraphics{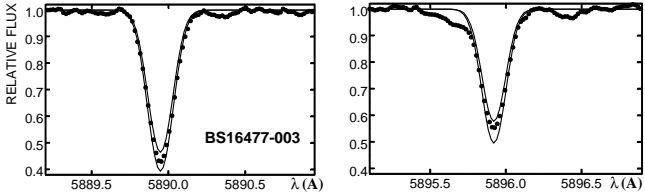}}
\caption[]{The observed spectra of the D1 and D2 lines ($dots$) in BS
16477-003 are compared to the theoretical profiles computed for the
best-fit Na abundance +0.10 and -- 0.10 ($solid~lines$).}
\label{Profil-b}
\end{figure}

\begin{figure}
\resizebox{\hsize}{!}{\includegraphics{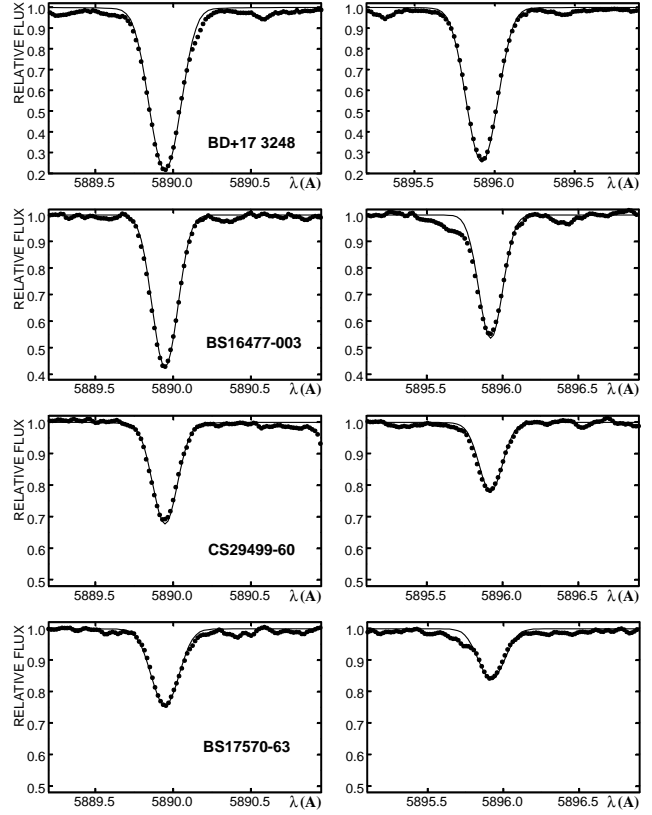}}
\caption[]{Best fit between observed ($dots$) and calculated
($solid~line$) profiles of the Na D1 (5889.95\AA) and D2 (5895.92\AA)
lines, for some program stars.}
\label{Profil-a}
\end{figure}

\section{Results}

New sodium abundances have been deduced from NLTE profiles of the D 
lines for our sample of EMP stars.
Typical uncertainty in the derived sodium abundance is about 
$\pm 0.05$ (see Fig. \ref{Profil-b}).
Figure  \ref{Profil-a} shows the best fitting between observed and
calculated profiles for some EMP giants and dwarfs.  In Table 1 
 the parameters of the models (Teff, log g and
${\rm v_{t}}$) which have been used and the new sodium abundance in
logarithm ${\rm \epsilon(Na)}$ (for ${\rm \epsilon(H)=12}$)
are given 
for each star. The
stellar ratios [Na/H] have
been computed assuming ${\rm \epsilon(Na)_{\odot}=6.25 }$ (see Sect.
\ref{analysis}).

Three stars in our sample are very peculiar carbon-rich stars:
CS~22892-52 (Sneden et al.  1996, 2000, 2003), CS~22949-37 (Norris et
al.  2001; Depagne et al.  2002), and CS~29528-41 (Sivarani et al.
2006).  The models we used for the NLTE computations do not take the carbon 
enhancement in these stars into account and thus the sodium
abundance computed here must be considered as very preliminary.  These
stars are not taken into account in the discussion.

\begin{figure}
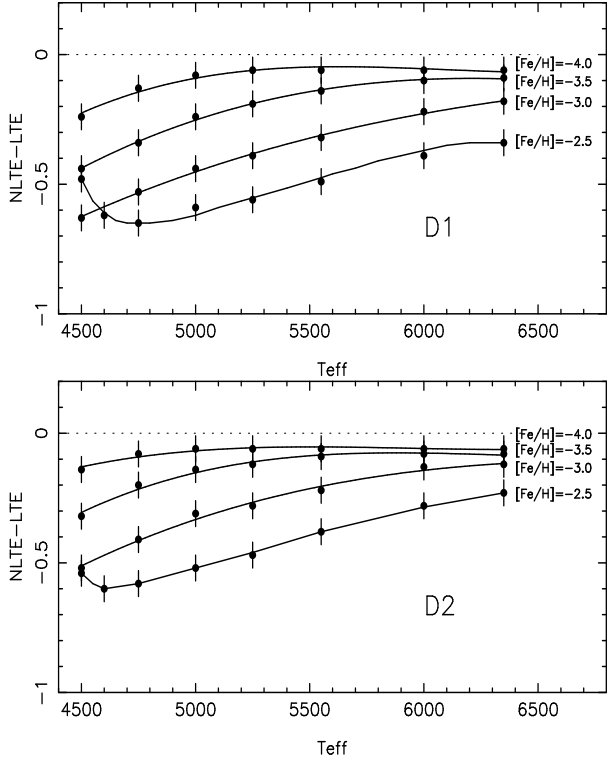

\resizebox{8.0cm}{5cm} 
{\includegraphics {6232fi5a.eps} }
\resizebox{8.0cm}{5cm} 
{\includegraphics {6232fi5b.eps} }
\caption[]{Individual correction applying to an LTE analysis
of the Na D1 and D2 lines along the RGB, for metal-poor stars 
with ${\rm -4.0<[Fe/H]<-2.5}$ and the hypothesis that [Na/Fe] = 0 .}
\label{corD}
\end{figure}

\begin{table}
\caption[]{Abundance corrections (${\rm \epsilon(NLTE)-\epsilon(LTE)}$) for 
the D1 and D2 lines.}
\label{cornlte}
\begin{tabular}{lc@{ }cc@{ }cc@{ }cc@{ }cc@{ }cc@{ }cc@{ }cc@{ }cc@{ 
    }cc@{ }}    
\hline 
[M/H] &  -2.5&&	      -3.0&&	        -3.5&&	          -4.0\\
\hline
\multicolumn{5}{c}{Teff=4500   log g=0.8}\\
   &   EW &  corr  &   EW &  corr	 &   EW &  corr    &  EW & corr\\   
D1 &  182 & -0.48  &  138 & -0.63	 &  107 & -0.44    &  81 &-0.24\\
D2 &  161 & -0.54  &  122 & -0.52	 &   93 & -0.32    &  66 &-0.14\\
\\
\multicolumn{5}{c}{Teff=4600   log g=1.0}\\  
D1 &  167 & -0.62  &	- &  -	         & -    &  -       &  -  & -\\ 
D2 &  149 & -0.60  &	- &  -	         & -    &  -       &  -  & -\\ 
\\
\multicolumn{5}{c}{Teff=4750   log g=1.4}\\ 
D1 &  151 & -0.65  &  124 & -0.53	 &  100 & -0.34    &  74 & -0.13\\
D2 &  135 & -0.58  &  110 & -0.41	 &   85 & -0.20    &  57 & -0.08\\
\\
\multicolumn{5}{c}{Teff=5000   log g=2.0}\\
D1 &  142 & -0.59  &  115 & -0.44	 &   91 & -0.24    &  63 & -0.08\\
D2 &  126 & -0.52  &  100 & -0.31	 &   75 & -0.14    &  45 & -0.06\\
\\
\multicolumn{5}{c}{Teff=5250   log g=2.7}\\ 
D1 &  137 & -0.56  &  108 & -0.39	 &   82 & -0.19    &  52 & -0.06\\
D2 &  119 & -0.47  &   92 & -0.28	 &   64 & -0.12    &  34 & -0.06\\
\\
\multicolumn{5}{c}{Teff=5550   log g=3.3}\\ 
D1 &  122 & -0.49  &   95 & -0.32	 &   67 & -0.14    &  38 & -0.06\\
D2 &  105 & -0.38  &   79 & -0.22	 &   49 & -0.09    &  23 & -0.06\\
\\
\multicolumn{5}{c}{Teff=6000   log g=3.7}\\ 
D1 &  101 & -0.39  &   75 & -0.22	 &   45 & -0.10    &  21 & -0.06\\
D2 &   85 & -0.28  &   57 & -0.13	 &   29 & -0.08    &  12 & -0.06\\
\\
\multicolumn{5}{c}{Teff=6350   log g=4.1}\\ 
D1 &   88 & -0.34  &   61 & -0.18	 &  32  &-0.09     &  13 & -0.06\\
D2 &   72 & -0.23  &   43 & -0.12	 &  19  &-0.08     &   7 & -0.06\\
\hline
\end{tabular}
\end{table}

In Fig.  \ref{corD} we present an estimate of the difference in sodium
abundance obtained when LTE or NLTE profiles of the D lines are used
along the Red Giant Branch of the HR diagram.  The computations have
been done for a set of temperatures between 6500~K (turn-off stars) to
4500~K (upper RGB stars) and metallicities from $-2.5$ to $-4.0$
(Table 2).  For each temperature the gravity has been deduced from the
isochrones of Kim et al.  (2002) for 14 Gyr, ${\rm [\alpha/Fe]=+0.6}$,
$\rm [Fe/H]=-3.7 and -2.7$ with the classical formula: $$
{\rm log~L/L_{\odot} = log~\mathscr{M/M}_{\odot} + 4~log~T_{eff} /
T_{eff,\odot} - log~g/g_{\odot}}
$$

\begin{figure}
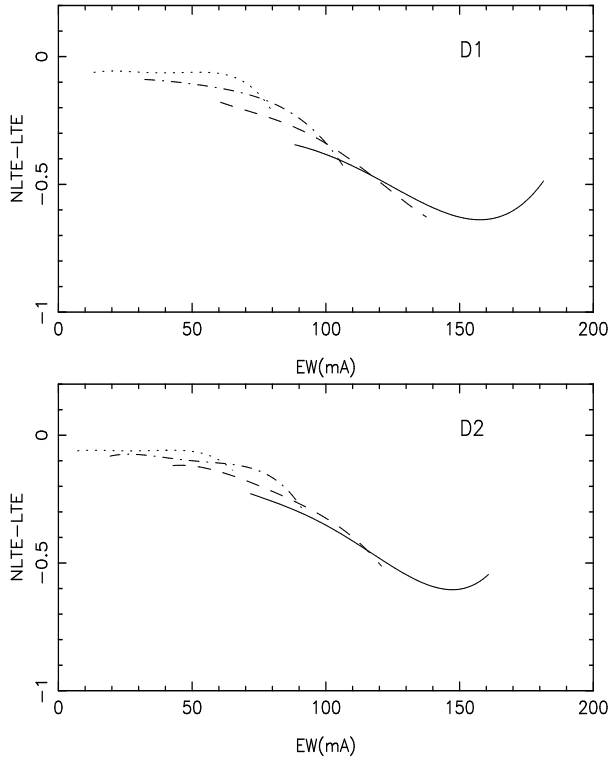

\resizebox{8.0cm}{5cm} 
{\includegraphics {6232fi6a.eps} }
\resizebox{8.0cm}{5cm} 
{\includegraphics {6232fi6b.eps} }
\caption[]{Individual corrections plotted as a function of the
equivalent widths of the lines for the different models given in Table
\ref{cornlte}.  The solid line represents the corrections for models
with [Fe/H]=-2.5, the dashed lines models with [Fe/H]=-3.0, the 
dotted-dashed lines models with [Fe/H]=-3.5, and the dotted lines models with
[Fe/H]=-4.0.  For a same equivalent width of the D lines the
correction little depends on the parameters of the models}
\label{wcorD}
\end{figure}

For this set of models, the LTE profiles have been computed with the
hypothesis that [Na/H] = [Fe/H] (i.e., [Na/Fe] = 0), and then fitted by
NLTE profiles that enabled the derivation of the appropriate NLTE
corrections.  A first cause of uncertainty is that the shapes of LTE
and NLTE profiles are slightly different, but the correction mainly
depends on the equivalent width of the sodium lines, as  can be seen
in Fig. \ref{wcorD}.
As a consequence, the corrections given in Table \ref{cornlte} can be
used only if the equivalent width of the lines is similar to that
given in the table.  If there is a strong anomaly of the sodium
abundance in the star, it is better to deduce the correction from the
Fig.  \ref{wcorD} (or to directly compute the non-LTE profile of the
line as it has been done for the stars of our sample).
Takeda et al.  (2003) have made a first attempt to compute the non-LTE
correction of the sodium abundance in a large metallicity range (${\rm
-4.0 <[Fe/H]<+0.4}$).  We have checked that in our domain of
metallicity (${\rm [Fe/H]<-2.5}$) there is a rather good agreement
with their computations.

\section{Discussion}

\subsection{Mixing effects in giant stars}

\begin{figure}
\begin {center}
\resizebox  {8.0cm}{5.0 cm} 
{\includegraphics{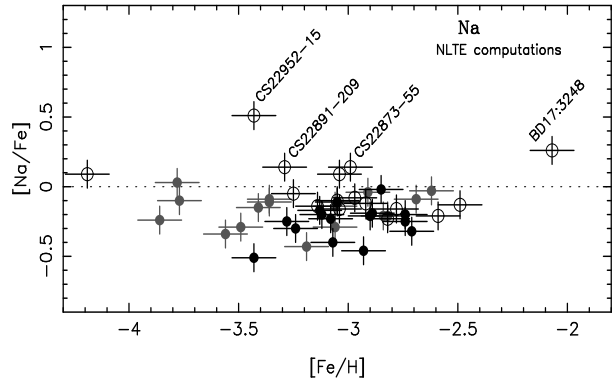}}
\caption[]{[Na/Fe] vs. [Fe/H] for the sample of EMP stars.  The symbols
are the same as in Fig.  \ref{na-lte}.  The sodium abundance has been
computed in NLTE with a modified version of the program of Carlsson.  The 
trend of the
[Na/Fe] ratio is flat in the region ${\rm -4.0<[Fe/H]<-2.5}$, and
there is a good agreement between the abundances of the giants and
turn-off stars.  All the stars with ${\rm [Na/Fe] > 0.1}$ have been
found to be extra-mixed, in Spite et al.  (\cite{SCP05}).
}
\label{na-nlte}
\end {center}
\end{figure}

In Fig.  \ref{na-nlte} we present the NLTE
values of [Na/Fe] for our sample of stars as a function of 
[Fe/H] (the carbon-rich stars have
been omitted).  The symbols are the same as in Fig \ref{na-lte}.
Dwarfs and giants now lie in the same locus, and the trend of the 
[Na/Fe] ratio, in the range ${\rm -4.0 <[Fe/H]<-2.5}$, is flat.  The mean value of
[Na/Fe] is close to --0.2.

A few stars appear to be "Na-rich" with ${\rm [Na/Fe] > +0.1}$.  All
these stars belong, following Spite et al.  (\cite{SCH06}), to the
class of the "mixed stars".  They present the characteristics of a
mixing with the H-burning shell, low abundance of carbon correlated
with a high abundance of nitrogen, and a low value of the ${\rm
^{12}C/^{13}C }$ ratio close to the equilibrium value.  However, all
the mixed stars are not Na-rich.  Let us remark that all the stars
found here with ${ \rm [Na/Fe] > +0.1}$ were already suspected from
LTE computations, to be Na-rich.  All of them, but BD 17:3248, have
also been found to be aluminum rich, and from their position in the HR
diagram they could be AGB stars (see also  Johnson
(\cite{Joh02}), and Bond (\cite{Bo80}) for BD 17:3248). None of the turn-off stars or
"unmixed" giants are sodium rich.  As a consequence the sodium rich
stars are suspected to have suffered a mixing between the atmosphere
and the H-burning shell,  a mixing deep enough to bring
the products of the Ne-Na cycle to the surface.
When only EMP dwarfs and unmixed giants are taken into account, we find
that ${ \rm [Na/Fe]= -0.21 \pm 0.13}$ or ${ \rm [Na/Mg]= -0.45 \pm
0.16}$; the star-to-star variation becomes small and comparable to what is 
obtained for the other elements (see Cayrel et al. \cite{CDS04}).

\subsection{Comparison with the nearby metal-poor stars, and the prediction 
of the models of galactic chemical evolution}

\begin{figure}
\begin {center}
\resizebox  {8.0cm}{5.0 cm} 
{\includegraphics{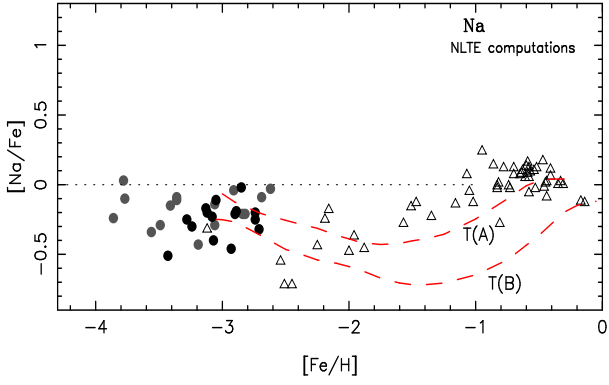}}
\caption[]{[Na/Fe] vs. [Fe/H] for the sample of EMP stars (Cayrel et
al. \cite{CDS04}; Spite et al.  \cite{SBC05} ), dwarfs (black dots),
unmixed giants (grey dots), and the nearby metal-poor stars studied by
Gehren et al.  (\cite{GSZ06}) (open triangles).  The dashed lines
represent the predictions of two models of Timmes et al.
(\cite{TWW95}) that differ by the quantity of iron ejected by massive
stars and thus by the position of the mass cut.}
\label{gehren-fe}
\end {center}
\end{figure}

Recently Gehren et al.  (\cite{GSZ06}) have computed the sodium
abundance in a sample of 55 nearby metal-poor stars, most of them
belonging to the thin or thick Galactic disc and some of them to the 
halo.  They have very carefully taken into
account  the non-LTE effects.  In Fig.  \ref{gehren-fe}
we compare the ratio [Na/Fe] in our EMP stars and in the sample of
Gehren et al. (\cite{GSZ06}); unfortunately, we have no star in common to check the
influence of systematic effects, but they have one star in our domain
of metallicity (G64-12 with [Fe/H]=--3.1), and this star falls exactly
in the middle of our sample in Fig.  \ref{gehren-fe}.

Sodium has only one (stable) isotope, ${\rm ^{23}Na}$, which is mainly
produced in the carbon burning process operating inside massive stars.
As noted by Woosley \& Weaver (1995) the resulting amount of sodium
nuclei in the SN ejecta is sensitive to the excess of neutrons,
therefore its production is somewhat metal dependent, but in the early
Galaxy (very low metal content) this element production should be the
same as for the primary elements.  There were several attempts to
incorporate the observed sodium abundance in the disc and halo stars
into Galactic evolutionary models.  One can mention here the works of
Timmes et al.  (1995), Samland (1998), and Goswami \& Prantzos (2000).
All the above-mentioned models are based on the Woosley \& Weaver
(1995) yields from massive stars, but use different initial-mass
functions, star formation rates, and other model specifications, which
can explain the slight differences in the predictions.

In Fig.  \ref{gehren-fe} our observations and those of Gehren et al.
(\cite{GSZ06}) are compared to the predictions of Timmes et al.
(\cite{TWW95}).  These authors predict a local minimum around [Fe/H] =
--1.5 with subsequent increasing of [Na/Fe] as the metallicity
decreases to [Fe/H] = --3.  The predictions for two different values
of the iron yields from massive stars (which depend on the mass cut in
the massive supernovae) are represented in the figure.  Between models
T(A) and T(B), the mass of iron ejected by SN~II differs by a factor
of two.  At a metallicity higher than [Fe/H]= --3, there is a rather
good agreement with the predictions of the model T(A), which supposes
the smallest value of the iron yields from massive stars, and such an
agreement is generally found for the other elements (see Cayrel et al.
\cite{CDS04}).

\begin{figure}
\begin {center}
\resizebox  {8.0cm}{5.0 cm} 
{\includegraphics{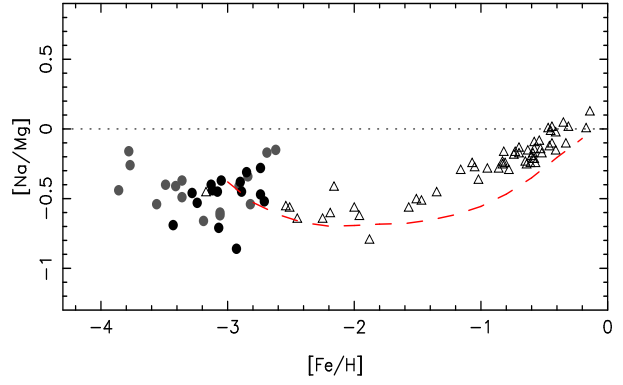}}
\caption[]{[Na/Mg] vs. [Fe/H] for the sample of EMP stars and the
nearby metal-poor stars studied by Gehren et al.  (\cite{GSZ06}).
Symbols are the same as in Fig.  \ref{gehren-fe}.  The dashed line
represents the prediction of the model of Timmes et al.
(\cite{TWW95}).  (The [Na/Mg] ratio is independent of the mass cut at
variance with [Na/Fe] and the predictions are reduced 
to a single 
curve).  It can
be seen in the figure that in the interval ${\rm -4 < [Fe/H] < -2.8}$
[Na/Mg] is constant.  However, a slight increase of [Na/Mg] when the
metallicity decreases (from [Na/Mg]= --0.5 at [Fe/H]=--3.  to [Na/Mg]=
--0.3 at [Fe/H]=--4.)  cannot be excluded.
}
\label{gehren-mg}
\end {center}
\end{figure}

In contrast with [Na/Fe], the ratio [Na/Mg] is independent of the mass
cut.  In Fig.  \ref{gehren-mg} we present [Na/Mg] as a function of
[Fe/H] for our sample of EMP stars and the sample of Gehren et al.
(\cite{GSZ06}).  (We neglected the NLTE effects on the magnesium lines
that are not supposed to be very important).  The agreement is rather
good for ${\rm [Fe/H]>-3}$.

Timmes et al. (\cite{TWW95}) do not predict the behaviour of [Na/Fe] 
or [Na/Mg] for ${\rm [Fe/H]<-3}$. From our measurements, 
[Na/Fe] and [Na/Mg] become flat: the relative abundances are 
independent of the metallicity.
However, a slight increase of [Na/Mg] when the metallicity decreases 
(from [Na/Mg]= --0.5 at [Fe/H]=--3. to [Na/Mg]= --0.3 at  [Fe/H]=--4.)  
cannot be excluded.

Samland (\cite{Sa98}) has considered the chemical evolution of many elements and
found that the expected distribution of [Na/Fe] vs.  [Fe/H] in the extreme
metallicity region from [Fe/H] = --4 to --2.5 is almost independent of
[Fe/H], and the mean [Na/Fe] value equals approximately --0.2.
This was not supported by observations available at that time, but is
in good agreement with our observations.  Qualitatively the same
conclusion was reached by Goswami \& Prantzos (2000).  According to
their model, [Na/Fe] must be almost constant within this metallicity
interval and equal approximately --0.5, but this is significantly 
lower than what is given by our calculations.  

Recently Tsujimoto et al. (\cite{TSY02}) have also investigated the chemical
evolution of sodium and aluminum during the early epochs of the
Galaxy.  They have used a model that reproduces the distribution of EMP
stars in the [Na/Mg] or [Al/Mg] versus [Mg/H] plane.  The observed
trends and the scatter of the observations are mainly based on the
work of McWilliam et al.  (\cite{MPS95}).  This sample contains
several mixed giants which explain the very large scatter of [Na/Mg]
and [Al/Mg] they observed at low metallicity.  Such a scatter, which is
an important point in the definition of the model of Tsujimoto et al.
(\cite{TSY02}), is not observed in the unmixed stars of our sample (dwarfs and 
"unmixed" giants) and thus does not probably reflect a scatter in the pristine material.
Let us note that Cohen et al.  (\cite{CCM04}), who studied a sample of
turnoff stars in the interval ${\rm -3.6 < [Fe/H] < -2.2}$, also found
a very small scatter of [Al/Mg].  The precise determination of [Al/Mg]
in our sample of stars from non-LTE synthetic profiles is in progress.

\section{Conclusion}

\indent

NLTE abundances of sodium have been
derived for a sample of extremely metal-poor dwarfs and giants 
using high-resolution spectra obtained with VLT. 
Both subsamples of dwarfs and unmixed giants now have the same mean 
[Na/Fe] or [Na/Mg] values, within the standard errors: ${ \rm 
[Na/Fe]= -0.21 \pm 0.13}$ or ${ \rm [Na/Mg]= -0.45 \pm 0.16}$.
However, a slight increase of [Na/Mg] at low metallicity, 
(from [Na/Mg]= --0.5 at [Fe/H]=--3. to [Na/Mg]= --0.3  [Fe/H]=--4.)  
cannot be excluded.

$\bullet$ In the domain  ${\rm  -4 < [Fe/H] < -2.5}$, 
the giants of our sample do not exhibit any systematic dependence of
[Na/Fe] upon [Fe/H], contrary to the preliminary findings of Cayrel et al.
(\cite{CDS04}). This effect was a consequence of a coarse treatment of 
the non-LTE effects: in Cayrel et al. (\cite{CDS04}) the LTE 
abundances had been uniformely corrected by --0.5dex. In fact, the 
non-LTE correction strongly depends on the equivalent width of the 
sodium lines and thus is generally smaller at lower metallicity. (When the 
lines are smaller they form deeper in the atmosphere where, in 
particular, collisions 
are more important and thus non-LTE corrections are smaller.) 

$\bullet$ Among the giant stars, the "mixed" giants (stars that have
undergone an extra-mixing between the atmosphere and the H-burning
shell following Spite et al.  \cite {SCP05}), often present an
overabundance of Na.  It is suggested that in these stars, material
processed in Ne-Na cycle in the hot hydrogen burning shell has been
brought to the surface.  From their position in the HR diagram, some of
these stars could be AGB stars.

$\bullet$ Qualitatively, our observational result on sodium
distribution in the metallicity region [Fe/H] from --4 to --2.5 agrees
with predictions of models elaborated by Samland (1998) and Goswami \&
Prantzos (2000).  Both models produce the almost constant [Na/Fe]
value in the region of extreme metal-poor stars.  In particular, the
Samland's model gives an average [Na/Fe] value that approximately equals
--0.2, and this value appears to be in quantitavive
agreement with our estimate of the mean relative-to-iron sodium
abundance in the early Galaxy.  The model of Timmes et al.  (\cite{TWW95})
also agrees with our observations around [Fe/H]=--3.  In qualitative 
agreement, the
model of Goswami \& Prantzos predicts a slightly too low relative sodium
abundance for extremely metal-poor stars.

\begin{acknowledgements}
SMA kindly acknowledges the C.N.R.S./INSU for financial support and the Paris-Meudon Observatoire for its hospitality during his productive stay in Meudon.
\end{acknowledgements}

\end{document}